\documentstyle[aaspp4,11pt,flushrt]{article}

\newcommand{\cmjj}{\mbox{${\rm cm^{-2}}$}}

\newcommand{\kms}{\mbox{km\ s${^{-1}}$}}
\newcommand{\lya}{\mbox{${\rm Ly}\alpha$}}

\begin{document}

\title{\Large\bf A High Deuterium Abundance at z=0.7$^1$}

\author{\large\bf
J. K. Webb\altaffilmark{2,3}, R. F. Carswell\altaffilmark{4},
K. M. Lanzetta\altaffilmark{5}, R. Ferlet\altaffilmark{6},
M. Lemoine\altaffilmark{7}, \\
A. Vidal-Madjar\altaffilmark{6}, D. V. Bowen\altaffilmark{8}
}

{\small \noindent $^1$ Based on observations with the NASA/ESA Hubble Space
Telescope, obtained at the Space Telescope Science Institute, which is operated
by the Association of Universities for Research in Astronomy, Inc., under NASA
contract NAS5--26555.}

{\small \noindent $^2$ School of Physics, University of New South Wales,
Sydney, NSW 2052, Australia}

{\small \noindent $^3$ Astronomy Centre, Sussex University, Falmer, Brighton
BN1 9QH, England}

{\small \noindent $^4$ Institute of Astronomy, Madingley Road, Cambridge, CB3
0HA, England}

{\small \noindent $^5$ Astronomy Program, Department of Earth and Space
Sciences, State University of New York at Stony Brook, Stony Brook, NY
11794--2100, USA}

{\small \noindent $^6$ Institut d'Astrophysique, 98 bis Boulevard Arago, Paris
75014, France}

{\small \noindent $^7$ Department of Astronomy \& Astrophysics, Enrico Fermi
Institute, The University of Chicago, Chicago IL 60637--1433, USA}

{\small \noindent $^8$ Royal Observatory Edinburgh, Blackford Hill, Edinburgh
EH9 3HJ, Scotland}

\vspace*{0.1in}
\hrule

{\bf Of the light elements, the primordial abundance of deuterium,
$({\rm D}/{\rm H})_p$, provides the most sensitive diagnostic$^1$ for the
cosmological mass density parameter $\Omega_B$.  Recent high redshift
${\rm D}/{\rm H}$ measurements are highly discrepant$^{2,3,4,5,6}$,
although this may reflect observational uncertainties$^{7,8}$. 
The larger ${\rm D}/{\rm
H}$ values, which imply a low $\Omega_B$ and require the Universe to be
dominated by non-baryonic matter (dynamical studies indicate
a higher total density parameter), cause problems for galactic
chemical evolution models since they have difficulty in reproducing
the large decline down to the lower present-day ${\rm D/H}$.
Conversely, low ${\rm D}/{\rm H}$ values imply an
$\Omega_B$ greater than derived from $^7$Li and $^4$He
abundance measurements, and may require a deuterium abundance
evolution that is too low to easily explain.  Here we report the first
measurement at intermediate redshift, where the observational
difficulties are smaller, of a gas cloud with ideal
characteristics for this experiment.  Our analysis of the $z = 0.7010$
absorber toward 1718$+$4807 indicates ${\rm D/H} = 2.0 \pm 0.5 \times
10^{-4}$ which is in the high range.  This and other
independent observations suggests there may be a cosmological
inhomogeneity in $({\rm D}/{\rm H})_p$ of at least a factor of ten.}

Measurements can be achieved using absorption line spectroscopy of gas
clouds intersecting the sight-lines to distant, background quasars by
measuring the strengths of the absorption lines of the H~I and D~I
Lyman series.  The D~I transitions occur at an isotopic shift of $-81$
km s$^{-1}$ with respect to the H~I series.  Such a measurement is
observationally difficult for a number of reasons$^{9,10}$: (1)
ill-placed H~I \lya-forest absorption lines can masquerade as
deuterium and (2) most QSO absorption systems exhibit complex internal
velocity structure, which may render parameter estimation for
individual components of interest unreliable or impossible.  To
overcome the first difficulty, we searched for suitable absorbers at
intermediate redshifts (i.e.\ at $z < 1$), where the number density of
\lya-forest systems is far lower than at high redshifts.  To overcome
the second difficulty, we searched for suitable absorbers with
apparently simple velocity structure.  We found one
object which appears optimal for a D/H analysis; only one absorbing
component is revealed by the data and the velocity dispersion in that
component is small enough to easily detect and accurately measure the
deuterium abundance.

The absorber at redshift $z = 0.7010$ toward the QSO 1718$+$4807
($z_{\rm em} = 1.084$, $m_V = 15.3$) was selected on the basis of a
remarkably abrupt partial Lyman-limit discontinuity in the
International Ultraviolet Explorer (IUE) spectrum$^{11}$.  The extreme
sharpness of the Lyman break clearly indicates simple velocity
structure {\it and} low velocity dispersion parameter.
This simple velocity structure contrasts with the complex nature
of the higher redshift systems reported so far, both those which give
low D/H measurements$^{2,3,7}$ and those which give high
values$^{4,5,6}$ so the measured column densities (based on
Ly$\alpha$ and the Lyman limit alone) are likely to be more reliable
than in those cases.  Hubble Space Telescope (HST) spectra were obtained of the
spectral region covering the Lyman-$\alpha$ and Si~III transitions
using the Goddard High Resolution Spectrograph and the G270M grating.
All spectra were extracted, binned to
linear wavelength scales and corrected to vacuum, heliocentric
wavelengths using standard procedures.
Absorption-line parameter estimates were derived using VPFIT$^{12}$, a
computer program based on Gauss-Newton unconstrained optimisation.
Parameter errors were derived from the diagonal terms of
the parameter covariance matrix computed at the best fit.  

The $\rm D/H$ we derive from the combined data is 
${\rm D/H} = 2.0 \pm 0.5 \times 10^{-4}$ which 
is approximately an order of magnitude 
larger than recent ground--based measurements of high-redshift
absorbers$^{2,3,7}$ (although the reliability of these measurements is
presently a matter of debate$^8$)
and an order of magnitude larger than a recent
intermediate redshift measurement$^{13}$ at z=0.5.  
It is of course possible that our high value is caused by a weak H~I line
which just happens to fall very close to the D~I wavelength.  Unfortunately,
{\it a posteriori} statistics do not provide a reliable means of
assessing the probability of this because the absorption cloud we have
studied was specifically selected as having uniquely simple velocity
structure, a property which itself implies a low probability of an H~I
interloper successfully mimicking D~I.  We note however that the
probability of a {\it randomly placed} interloper mimicking deuterium
is extremely small; using the observed number density of absorbers in
the spectrum and adopting a generous 4-sigma tolerance on the line
position, it is less than 1 percent.

To examine this possibility further, we re-fitted the data replacing
D~I with H~I (so that the redshift of the putative interloper was a
free parameter).  The best fit resulted in the new H~I line falling at
$-86 \pm 5$ \kms\ from the strong component, implying that the
detected absorption feature {\it is} deuterium and not an interloper.
Furthermore, the Doppler parameter of this putative H~I interloper ($b
= 21 \pm 4$ \kms) lies between that of the strong H~I component but
greater than the Si~III value, which is again consistent with the
expectation for D I.  Taken together, these points indicate that the
most reasonable and likely interpretation of the data is that we have
detected D~I.

A further possibility is that some other transition from an absorbing
cloud at some other redshift just happens to coincide with the D~I
wavelength at $z_{\rm abs} = 0.701024$.  The closest candidate is ISM
Cr II $\lambda 2066.16$.  The D~I feature falls at 2067.31 \AA.  This
corresponds to a shift of $\sim170~ \kms$, thus so Cr~II
$\lambda 2066.16$ is not a serious contender for D~I contamination.
Furthermore, Cr~II $\lambda 2056$ is not observed and yet has an
oscillator strength 2 times higher than Cr~II $\lambda 2066.16$.  We
conclude that no Cr~II contamination occurs.

We cannot reliably estimate the heavy element abundances of the
$z_{\rm abs} = 0.701024$ system, since the only species in the
observed range is Si~III ($\log N({\rm Si~III}) = 12.81 \pm 0.04$
\cmjj).  An upper limit can be derived, but this only constrains the
abundances to be less than solar; they could easily be very much
lower.  We note that it would be astonishing if the $z_{\rm abs} =
0.701024$ system turned out to have already undergone substantial
chemical evolution and yet exhibit such a high D/H.  Future
observations of any unsaturated absorption lines in this system
(potentially Mg~II or Fe~II, for example) should yield reliable heavy
element abundance constraints.  Astration (deuterium destruction in
stars) means that any particular D/H detection provides a {\it lower}
limit to the primordial value.  Therefore, because our measured D/H is
high, the lack of information about the heavy element abundances does
not alter the interpretation of the data described below.

In standard models of galactic chemical evolution, the mass fraction
of deuterium decreases steadily with time, by a factor $\sim2-3$ over
10 Gyr, so that when plotted versus metallicity, it appears constant
until a metallicity 1/10 solar and drops abruptly thereafter$^{14}$.
The two extremes of the recent high redshift D/H observations imply
radically
different consequences for the baryonic density parameter, $\Omega_B$,
and its cosmological significance$^{15,16}$.  A high $({\rm D}/{\rm
H})_p$ would be in excellent agreement with standard homogeneous BBN
and the observed ``primordial'' abundances of $^4$He and $^7$Li, but
observations of D/H in the interstellar medium then require
destruction of deuterium by a factor $\sim10$ up to the present epoch.
Whether this much astration can be easily explained by chemical evolution
models is unclear, although some recent models may succeed$^{17}$.

A ratio ${\rm D}/{\rm H}=2.0 \times 10^{-4}$, as obtained here toward
Q1718+4807, corresponds to a mass fraction of $^4$He $Y \simeq 0.233
\pm 0.002$, $^7{\rm Li}/{\rm H} \simeq 1.8^{0.7}_{0.6} \times
10^{-10}$ and $\Omega_B \simeq 0.006 \pm 0.003 \ h^{-2}$, where $h$
denotes the value of the Hubble constant in units of 100 km s$^{-1}$
Mpc$^{-1}$ and the errors only include $1 \sigma$ errors on the
nuclear cross-sections of BBN (values were derived using useful
fitting formulae$^{18}$).  Remarkably, these figures agree precisely
with the measured abundances, $Y \simeq 0.233 \pm 0.005$ and $^7{\rm
Li}/{\rm H} \simeq 1.8^{+0.5}_{-0.3} \times 10^{-10}$. The above
baryonic mass density parameter may be compared with the measured
amount of visible matter in galaxies, $\Omega_{\rm vis} \sim
0.002-0.005 \ h^{-1}$. This shows that a high primordial deuterium
abundance leaves little room for {\it baryonic} dark matter and
supports the view that the missing mass inferred from dynamical
studies must be non-baryonic.  On the other hand, a low $({\rm D}/
{\rm H})_p$ implies BBN values of $^4$He and $^7{\rm Li}/{\rm H}$
which differ from the measured values by amounts corresponding to $2
\sigma$ observational limits (although it avoids the difficulty in
explaining such rapid deuterium destruction, but if too low, may still
be problematic if the implied astration is too small compared to
predictions).

Several explanations have been put forward which could explain
discrepant values of the D/H ratio.  Post-BBN chemical evolution
processes might result in strong deuterium depletion in some gas
clouds, giving rise to low observed abundances.  By redshifts $z
\approx 3-4$, stars down to $\sim 2 M_{\odot}$ have had time to eject
deuterium-poor gas.  However, the observational upper limits on C/H
and Si/H in clouds at high redshift where ${\rm D/H} \sim 2.5 \times
10^{-5}$ imply metal abundances less than $\sim 10^{-2}$ solar,
constraining the fraction of gas which has been cycled through stars
to be less than about 5\%.  Such models appear to offer reasonable
compatibility with light element abundance observations only by
admitting unreasonable stellar populations, such as a primordial
population of supermassive stars of mass $M \gtrsim 1000M_{\odot}$, or
an initial mass function strongly peaked around $M\approx 6
M_{\odot}$.

Alternatively, primordial isocurvature baryon fluctuations$^{19}$
could account for a variation of the $({\rm D}/{\rm H})_p$ ratio by a
factor $\sim 10$ but only on scales corresponding to a Jeans mass of
$M_J \sim 10^5-10^6 M_{\odot}$.
For a specific class of these models, there are opposing views as to
whether or not the observed isotropy of the cosmic microwave
background rules out substantial {\it intrinsic} D/H
fluctuations$^{19-22}$.  We note, however, that homogeneous,
critical-density models with a low baryonic component, $\Omega_B <
0.01~h^{-2}$, appear to predict degree--scale microwave background
fluctuations which are significantly below those actually
observed$^{23}$.  Also, assuming no segregation between baryons and
dark matter$^{24}$, X-ray observations of galaxy clusters support this
point$^{23}$ since for critical-density models they independently
suggest baryonic density parameters of $\Omega_B > 0.02~h^{-3/2}$.
Both of these types of observation therefore appear to be in conflict
with a baryonic density parameter for a homogeneous universe as low as
that derived from our results for D/H in the $z_{\rm abs} = 0.701024$
gas cloud reported here.  Therefore our results indicate that either
the universe does not have a critical total density, or, if it does,
Big-Bang nucleosynthesis must have occurred inhomogeneously.

\vspace*{0.1in}
\hrule

\begin{small}

\noindent 1. Epstein, R.I., Lattimer, J.M., Schramm, D.N., 1976,
Nature, 263, 198-202.
The Origin of Deuterium.

\noindent 2. Tytler, D., Fan, X-M., Burles, S., 1996, Nature, 381, 207-209.
Cosmological baryon density derived from the deuterium abundance at
redshift z = 3.57.

\noindent 3. Burles, S., Tytler, D., 1996, Science, submitted.
Cosmological deuterium abundance and the baryon density of the
universe.

\noindent 4. Rugers, M., \& Hogan, C.J., 1996, {\it Astrophys. J.}, 
469, L1-L4.
Confirmation of High Deuterium Abundance in Quasar Absorbers

\noindent 5. Carswell, Rauch, Weymann, Cooke, Webb, 1994, 
{\it Mon. Not. R. Astron. Soc.}, 268, L1-L4.
Is there deuterium in the z = 3.32 complex in the spectrum of 0014+813?

\noindent 6. Rugers, M., Hogan, C.J., 1996, {\it Astron. J.}, 
111(6), 2135-2140.
High Deuterium Abundance in a New Quasar Absorber

\noindent 7. Tytler, D., Burles, S., Kirkman, D., 1996, preprint
astro-ph/9612121.
New Keck spectra of Q0014+813: annulling the case for high
deuterium abundances.

\noindent 8. Songaila, A., Wampler, E.J., Cowie, L.L, 1997, 
{\it Nature}, 385, 137-139.
A high deuterium abundance in the early Universe.

\noindent 9. Webb, J.K., Carswell, R.F., Irwin, M.J., Penston, M.V.,
{\it Mon. Not. R. Astron. Soc.}, {\bf 250}, 657-665, 1991.
On measuring the deuterium abundance in QSO absorption systems

\noindent 10. Wampler, E.J., {\it Nature}, {\bf 383}, 308, 1996.
Alternative hydrogen cloud models.

\noindent 11. Lanzetta, K.M., Turnshek, D., Sandoval, J., 1993, 
{\it Astrophys. J. Suppl.}, 84, 109-184.
Ultraviolet spectra of QSOs, BL Lacertae objects, and Seyfert galaxies

\noindent 12. Webb, J.K, PhD thesis, Cambridge University, 1987.
QSO absorption lines.

\noindent 13. Vidal-Madjar, A., Ferlet, R., Lemoine, M, 1996, preprint
astro-ph/9612020.
Deuterium abundance and cosmology.

\noindent 14. Pagel, B.E.J., {\it Nucleosynthesis and Chemical Evolution
of Galaxies}, Cambridge University Press 1997.

\noindent 15. Schramm, D.S., Turner, M.S., {\it Nature}, 1996, 381, 193-194.
Deuteronomy and numbers.

\noindent 16. Hata, N., Steigman, G., Bludman, S., Langacker, P., 1996,
preprint astro-ph/9603087.  Cosmological implications of two
conflicting deuterium abundances.

\noindent 17. Vangioni-Flam, E., Cass\'e, M., 1995, {\it Astrophys. J.} 
{\bf 441}, 471-476.
Cosmological and astrophysical consequences of a high
primordial deuterium abundance

\noindent 18. Hogan, C.J., preprint, astro-ph/9609138.
Big bang nucleosynthesis and the observed abundances of light elements. 

\noindent 19. Copi, C.J., Olive, K.A., Schramm, D.N., 1996, preprint
astro-ph/9606156.  Implications of a primordial origin for the
dispersion in D/H in quasar absorption systems.

\noindent 20. Copi, C.J., Schramm, D.N., Turner, M.S., 1995, 
{\it Science}, {\bf 267}, 192.
Big-Bang Nucleosynthesis and the Baryon Density of the Universe.

\noindent 21. Jedamzik, K., Fuller, G.M., 1995, {\it Astrophys. J.}, 
{\bf 452}, 33.
Nucleosynthesis in the Presence of Primordial Isocurvature Baryon
Fluctuations.

\noindent 22. Jedamzik, K., Fuller, G.M., 1996, preprint astro-ph/9609103.
Is the deuterium in high redshift Lyman limit systems primordial?

\noindent 23. White, M., Viana, P.T.P., Liddle, A.R., Scott, D., 1996,
{\it Mon. Not. R. Astron. Soc.}, {\bf 283}, 107.
Cold dark matter models with high baryon content

\noindent 24. White, S.D.M., Navaro, J.F., Evrard, A.E., Frenk, C.S., 1993,
{\it Nature}, {\bf 366}, 429.
The Baryon Content of Galaxy Clusters:  A Challenge to Cosmological
Orthodoxy.

\end{small}
\vspace*{0.1in}
\begin{small}
\noindent
ACKNOWLEDGEMENTS. KML acknowledges support by NASA, STScI, and NSF.  ML
acknowledges support by the NASA, DoE, and NSF at the University of Chicago.
JKW is grateful for valuable discussions with M. Ashley, J. Barrow, A. Liddle, 
and R.J. Tayler and to SUN Microsystems Australia Pty Ltd for providing 
computing facilities for this work.
\end{small}

\newpage

\figcaption{Data and fits for 1718$+$4807.  The top panel illustrates 
the HST H~I and D~I profile, showing the model profiles for H~I (dotted 
line), D~I (dashed line) and H~I+D~I (solid
line).  The middle panel illustrates the HST Si~III $\lambda 1206.51$
profile and fit.  The HST spectrum was observed on 3/3/1995
(total integration time 282 minutes).  The spectral resolution
is 20,000.  The bottom panel illustrates the IUE spectrum and fit.  
The pixel size is 1.18\AA~ and the adopted spectral
resolution is 2.5 pixels.   To minimise the number of free parameters, 
we adopted a single redshift for H~I, D~I and Si~III lines and the 
Lyman limit.  The parameter constraints arise as follows.  The IUE 
spectrum provides an accurate
determination of the H~I column density ($\log N({\rm H~I}) = 17.24
\pm 0.01$ \cmjj).  The Si~III $\lambda$1206 absorption line supports
the single component velocity structure implied by the IUE Lyman limit
and determines the cloud redshift precisely ($z_{\rm abs} = 0.701024
\pm 0.000007$).  The \lya\ absorption is clearly asymmetric, 
showing additional absorption in the
blue wing at the position corresponding to D~I.  Since
Si~III accurately constrains the D~I position and
because the Doppler parameter (which is equal to $\sqrt 2$ times the
RMS velocity dispersion) $b({\rm D~I})$ is constrained by the
overall fit, only one free parameter, $N({\rm D~I})$, is
required to fit the excess absorption seen at the D position.
The D~I column density is thus accurately
determined and is $\log N({\rm D~I}) = 13.57 \pm 0.06$ \cmjj.  The
Doppler parameters are dominated by non-thermal broadening with an
inferred temperature of $1.9 \times 10^4$ K and
$b({\rm H~I}) = 25.5$ \kms\ and $b({\rm D~I}) = 22.2$ \kms\ and 
$b({\rm Si~III}) = 18.7$ \kms, with the same error on each
of $\sigma = 0.5$ \kms.  From the above, we derive
${\rm D/H} = 2.0 \pm 0.5 \times 10^{-4}$.}

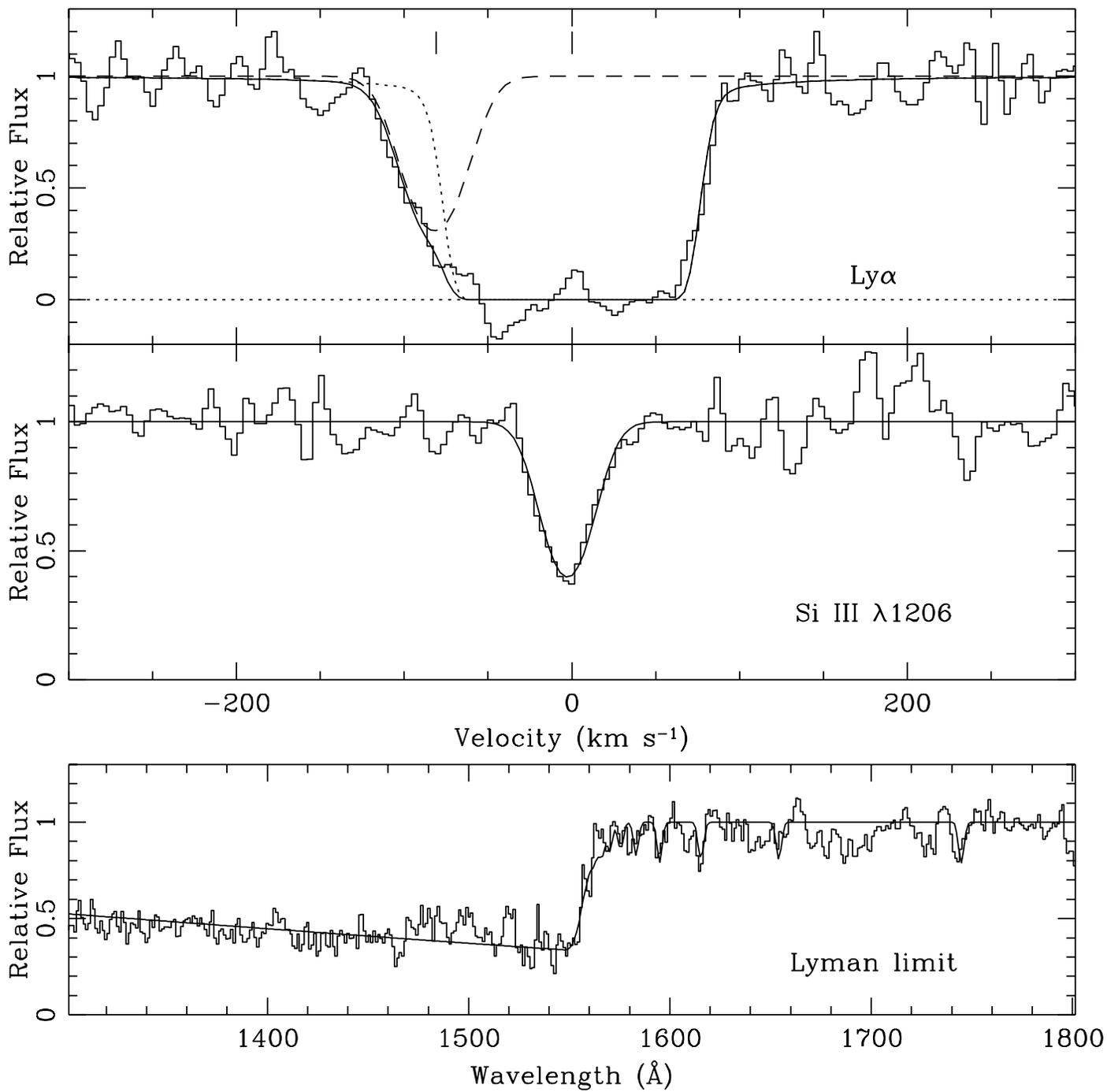
\end{document}